\documentstyle[12pt,epsf,aasms]{article}
\slugcomment{Submitted to Astrophysical Journal Letters, Feb 5, 1994}
\newcommand{\etal}{et~al.}
\def \paczy{Paczy{\'n}ski}
\begin{document}
\title{Gamma-Rays from Grazing Incidence Cosmic Rays
in the Earth's Atmosphere}
\author{Andrew Ulmer\altaffilmark{1}}
\affil{Princeton University Observatory,
       Peyton Hall,
       Princeton, NJ~08544--1001,
       USA}
\altaffiltext{1}{E-mail: andrew@astro.princeton.edu}

\begin{abstract}

Interactions of grazing incidence,
ultra high energy cosmic rays with the earth's atmosphere
may provide a new method of studying energetic cosmic rays with gamma-ray
satellites.
It is found that these cosmic ray interactions may produce gamma-rays
on millisecond time scales which may be detectable by satellites.
An extremely low gamma-ray background for transient gamma-ray events
and a large area of interaction,
the earth's surface, make the scheme plausible.
The effective cross section of detection of interactions for
cosmic rays above $10^{20}$~eV is found to
be more than two orders of magnitude higher than earth based detection
techniques.
This method may eventually offer an efficient way of
probing this region of the cosmic ray energy spectrum where
events are scarce.
In this paper, a conceptual model is presented for
the production of short bursts of gamma-rays based on these grazing incidence
encounters with the earth's atmosphere.

\end{abstract}


Knowledge of the cosmic ray spectrum above $10^{20}$~eV would provide important
insight into both the acceleration and propagation of ultra high
energy cosmic rays (UHE's). For instance, it has long been believed that there
is a cutoff in the energy spectrum due to interactions between the
cosmic rays and the microwave background (\cite{gr}; \cite{zk}).
Despite recent progress (cf. \cite{Bird}), and work over the last three
decades,
the study of this energy range remains limited by
statistics and instrumental effects of earth based cascade detectors.
For these reasons, it is of cardinal importance to consider
alternative techniques to extend the study of this region.
It is the purpose of this work is to investigate
the feasibility of a detection method based on grazing incidence cosmic
rays in the earth's atmosphere.
A conceptual model of the gamma-ray production
as well as some of the detectable characteristics are discussed.
Specifically, energy requirements, frequency of
occurrence, timescale, and energy spectrum are estimated.
Parenthetically, this study was motivated by the announcement of
the recent detection of what are believed
to be short gamma-ray flashes from the direction of the earth by
the {\it Burst and Transient Source Experiment} (BATSE) on the
{\it Compton Gamma Ray Observatory} (CGRO) at the 1993 Huntsville Gamma-ray
Burst Conference (\cite{fish}), but no attempt is made here to
make a detailed comparison with the reported phenomenon. It cannot yet be
determined if there is a link between the grazing incidence cosmic rays
and the detected flashes; however, UHE's are important
regardless of their connection, or lack therof, to the BATSE events.

Cosmic rays are known to be incident on the earth at
energies up to and above $10^{20}$~eV.
Ultra high energy cosmic rays interact with the
atmosphere causing a cascade of particles.
Most observations of these cascades are ground based. However, it has
been suggested that cosmic rays of $10^{19}$~eV and above could
be studied from space by observing optical ionization emission in the
atmosphere (\cite{BRA}, \cite{Li}), although these methods were finally
found to be impractical due primarily to the large optical background.
Additionally, interactions on the
solar surface of lower energy cosmic rays have previously
been investigated theoretically and are suspected to produce a few
$\sim 100$~MeV gamma-rays at a marginally detectable level (\cite{SSG}).
Most relevant to the question of cosmic ray induced gamma-rays,
is the earth gamma-ray background. This well-studied background is believed
to be produced predominately by the bremsstrahlung of electrons
from the cascades of cosmic rays (e.g. \cite{Be}; \cite{INR}; \cite{De}).
The detection
method proposed here
is based on intensity fluctuations in time of this background produced
by the most energetic cosmic rays.

The reason for interest in grazing as opposed to oblique
incidence cosmic rays is the atmosphere
is highly opaque to gamma-rays, with an attenuation depth
of $\sim$15 gm/cm$^2$ (e.g. \cite{M}). Cascades which produce gamma-rays
deep in the atmosphere will not be seen by satellites.
Gamma rays at an altitude of $\sim$30 km will travel
one attenuation length, therefore,
any atmospheric gamma-rays detectable from space
are most likely to be generated in the stratosphere (25-50 km).
In a typical cascade, the number of particles will increase
while degrading in energy via strong interactions, bremsstrahlung,
and pair production.
When most of the constituents of the cascade have
energies of $\sim$80~MeV, the cascade elements are primarily electrons,
positrons, and photons. At this energy, the process reaches a maximum
number of particles after which the electrons quickly lose their
energy to ionization (cf. \cite{S}, \cite{Lo},\cite{G}).
It takes UHE's about 800 gm/cm$^2$ to
reach this critical energy, whereas the column density at sea level is about
1000 gm/cm$^2$.  At maximum, the cascade is almost exclusively electrons
and positrons and contains nearly all of the energy of the initial
particle (\cite{G}).
If the particles move tangent to the earth's surface and
are high enough in the atmosphere so that the density is much smaller
than at sea level, the entire cascade can have a path length of the
order of hundreds of kilometers.  If beamed similarly to downward moving
particle cascades,
the lower energy electrons are beamed into a cone of half-angle
approximately 4$^{\circ}$ (e.g. \cite{Bird}).
Additionally, at energies near the
critical energy where the cascade has a maximum number of charged particles,
the electrons will begin to spiral in the earth's
magnetic field with radii of less than a kilometer which is much shorter than
the path length.
Although the radiation,
which is primarily bremsstrahlung,
from a single one of these electrons is beamed
along the direction of motion,
the gyration of the electrons will randomize
the directions of motion and emission into a plane.
If the incident cosmic ray is perpendicular to the magnetic field,
the cascaded radiation will be beamed into approximately 0.9 sterradians
corresponding to the angular area of a rotated cone of 4$^{\circ}$.
However, if the incident direction is parallel, the magnetic field
will not have a large effect on the beaming area. For the calculations in the
rest of the paper, an average beaming area of 0.5 sterradians is assumed.
Planer randomization
is not as important in the usual downward moving cascades detected from earth
(e.g. \cite{elbert})
because the total path length of the subcritical energy
particles is much shorter. If the cascade products are high
enough in the atmosphere ($\sim$15 gm/cm$^2$), some of the
gamma-ray photons will escape and may be detectable.

Currently, the most sensitive all-sky gamma-ray detector,
BATSE is capable of detecting and directionally
localizing a burst with about one hundred photons in the 50--300~keV
range (\cite{Fia}).
With this as a typical energy required for a detection, the total
energy requirement
in the initial particle can be estimated. If the radiation is beamed as
discussed above and emitted from the stratosphere at a distance of 500
kilometers from the detector,
then the total energy needed to detected by BATSE with an area
of $\sim$1 m$^2$ is estimated to be
\begin{equation}
(0.5 ster rad)(500 km)^2(300 keV)(100 photons)/\epsilon_{-2} \approx 4\times
10^{20} eV/\epsilon_{-2},
\end{equation}
where $\epsilon_{-2}$ is taken as an efficiency normalized to 1 per cent.
If the efficiency of converting energy into
gamma-rays is approximately 1 per cent,
then the method would examine extremely ultra high energy
events, although in the future, larger, more sensitive experiments may be
capable of observing the spectrum at lower energies.
Although the efficiency
should eventually be derived from extensive monte carlo simulations, it
can be estimated by taking the ratio of the energy flux of
the upward atmospheric gamma-ray background (\cite{INR}) to the energy flux
of gamma-ray producing incident cosmic rays.
Such estimates give values ranging from $\epsilon_{-2} = 0.2-1$.

The integral flux of cosmic rays above $4\times$10$^{20}$~eV
is estimated to be $\sim 1.5\times 10^{-20}$~cm$^{-2}$s$^{-1}$ ster~rad$^{-1}$
by extending the spectrum from lower energies
(e.g. \cite{Bird}; \cite{S}). The area of the atmosphere
observable by a satellite can be calculated to be
approximately
\begin{equation}
2\pi rR \approx 1\times 10^7 km^2
\end{equation}
where $R$, the radius of the earth, is 6000 km and $r$, the height of a
satellite orbit (taking GRO for example), is 400 km.
The fraction of acceptable incidence angles to the
earth's surface can be
estimated. The cascades travel many hundreds of kilometers in
the atmosphere, and assuming it must reach maximum
in about a 10 kilometer range of altitudes, incident angles to the horizon are
required to be within a range of about $\pi/100$.
The second angular component is arbitrary so, the
amount of acceptable incidence angles is $\sim$$0.2$ sterradians
Therefore, assuming this area of detection,
acceptable incidence angles, and a beaming factor of individual events of
$(0.5/4\pi)$ the total number of detectable
events in a one year period is found to be
\begin{equation}
\frac{0.5}{4\pi}(1.5\times 10^{-20})(1\times 10^{7}km^2)(3\times 10^7sec)(0.2
ster rad) \approx 350~ events.
\end{equation}
Because of the steep slope of the cosmic ray spectrum,
the number of events can be increased/decreased by a factor of 100 for
every decade decrease/increase of input cosmic ray energy above about
10$^{16}$~eV.
Consequently, the frequency of expected events scales strongly with
the cosmic ray energy necessary for detection, and therefore $\epsilon_{-2}$.

The timescale of cosmic ray events are likely to be no more than
a few milliseconds, because the light travel time across the earth
is about 40 milliseconds, and only a fraction of the earth is observed by
most satellites. This timescale is comparable to
the timescales of a few milliseconds
that have been observed by BATSE in the perturbation of arrival
times of a short gamma-ray burst by reflection off the earth's
atmosphere (\cite{Bh}).
For relativistic particles, two milliseconds corresponds to
distance scale of 600 km.  This distance is consistent
with the region of atmosphere observed by a satellite in orbit at a
height of 400 km.  Such a path length is plausible for a cascade in
the atmosphere because the depth of shower maximum which
occurs when the electrons reach the critical energy is nearly 1000
gm/cm$^2$ for the most energetic UHE's, and the density at a height of
30 kilometers
is $2\times 10^{-5}$ g/cm$^3$ (\cite{Usatm}).
Therefore, a cascade which lasts hundreds of km is possible.

Similar to the atmospheric background of gamma-rays, the UHE cascades
may produce gamma-rays
by bremsstrahlung or pair-annihilation (\cite{Be}; \cite{ling}).
A rough estimate of the spectrum of grazing incidence
UHE events is the background
gamma-ray spectrum of the earth's atmosphere as observed by
satellite. This background is believed to be primarily due to
lower energy cosmic ray interactions and is characterized
by a power law energy spectrum of
E$^{-1.4}$~counts/cm$^2$/s/keV around 1~MeV (\cite{INR}, \cite{De}).

In conclusion, it appears that it is within the capabilities of current
or next generation gamma-ray experiments to search for
grazing incidence ultra high energy cosmic rays as short bursts of gamma-rays
in
the earth's atmosphere.
One possible signature of UHE interactions is that,
because some secondary products of the UHE may not degrade immediately,
a few different maxima, or gamma-ray emission locations, may be
produced which would differ by a distance of
hundreds of kilometers.
In these cases, the bursts may not appear as point sources to
BATSE which may be able to localize the directions to better than 100
kilometers although the localization of bursts is a strong function of the
incident spectrum and may be worse for spectra which differ from those of
classical gamma-ray bursts for which the experiment was optimized
(\cite {Brock}; \cite {Ho}).
However, if many events were detected, it may be possible to
statistically argue that the events are not point sources.
Additionally, a detailed spatial distribution function could show
anisotropies due to the earth's magnetic field or column densities
in the atmosphere.
There are many geometries and possible interactions that
Can conceivably produce a large variety of events.
For this reason, quantitative monte carlo
studies of these interactions would yield more exact predictions of what
would be observed from space.
These ultra high energy cosmic ray interactions
may eventually reveal valuable information about UHEs.
For example, contingent on the assumptions presented above,
if BATSE has detected none of these UHE events in two years of
operation, one can place an upper limit on the integral flux
of cosmic rays
above approximately $4\times 10^{20}$~eV
of ~$\sim 2 \times 10^{-22}$~cm$^{-2}$s$^{-1}$ ster~rad$^{-1}$.

\acknowledgements

It is a pleasure to thank B. \paczy, M. Ulmer, G. Fishman, T. Stanev, and the
anonymous referee for helpful comments.
This project was supported by the NASA grant NAG5-1901.

\vfill\eject
\end{document}